\documentstyle[12pt]{article}       
\begin{document}    
\baselineskip=24.5pt    
\setcounter{page}{1}         
\topskip 0 cm    
\begin{flushright}    
{\large UT-788}\\
{hep-ph/9709388}\\
\end{flushright}    
\vspace{1 cm}    
\centerline{\Large \bf Atmospheric Neutrino Oscillation and}
\centerline{\Large \bf a Phenomenological  Lepton Mass Matrix}
\vskip 1 cm
\centerline{\large M. Fukugita$^{1,2}$, M. Tanimoto$^3$ and T. Yanagida$^4$}
\vskip5mm
\centerline{$^1$ Institute for Cosmic Ray Research, University of Tokyo,
Tanashi, Tokyo 188, Japan}
\centerline{$^2$ Institute for Advanced Study, Princeton, NJ 08540, U. S. A.}
\centerline{$^3$ Faculty of Education, Ehime University, Matsuyama 790-77,
Japan}
\centerline{$^4$ Department of Physics, University of Tokyo, Tokyo 113, Japan}
\vskip 2.5 cm
\noindent
{\large\bf Abstract}
\medskip

\noindent
We propose simple phenomenological lepton mass matrices which 
describe the three neutrinos almost degenerate in mass, leading
to a very large mixing angle between $\nu_\mu$ and
$\nu_\tau$, as consistent with a recent report on atmospheric
neutrino oscillation from the Superkamiokande collaboration. Our
matrix model also gives $\nu_e-\nu_\mu$ mixing in agreement with
the value required for neutrino oscillation to explain the solar
neutrino problem.

\newpage 
\noindent
A recent report on the atmospheric neutrino from the Superkamiokande 
collaboration [1] has presented convincing evidence that the long-standing
problem of muon neutrino deficit in underground detectors [2] is 
indeed due to neutrino oscillation.  The most surprising feature 
for theorists is a very large
mixing angle close to maximal between $\nu_\mu$ and 
its oscillating partner in contrast to the quark sector for which mixing
among different generations is all small.  This points towards the lepton 
mass matrix being governed by a
rule significantly different from the one that is relevant in the quark sector.

Accepting this atmospheric neutrino result from Superkamiokande and assuming
also that solar neutrino problem is ascribed to neutrino oscillation
(either matter enhanced [3] or usual oscillation in vacuum [4]), 
we may think of
two distinct possibilities for the neutrino mass, i.e., (i)
hierarchical massive neutrinos,
$$m_{\nu_e}\ll m_{\nu_\mu}\ll m_{\nu_\tau}, \eqno{(1)}$$
or, (ii) almost degenerate massive neutrinos
$$m_{\nu_e}\approx m_{\nu_\mu}\approx m_{\nu_\tau}, \eqno{(2)}$$
where in both cases  the $\nu_\mu-\nu_e$ mass difference is prescribed 
by oscillation for
solar neutrinos, and $\nu_\tau-\nu_\mu$ by the atmospheric neutrino
oscillation experiment.

In this letter we explore the possibility whether the experimentally
indicated lepton masses and mixings can be derived from a 
lepton mass matrix that is consistent with some simple symmetry
principle, hopefully, as much parallel as possible to that for the
quark sector. The problem of quark-lepton mass is one of the most
difficult problems in particle physics, and we have no theory 
that predicts the mass matrix from a known principle. The best 
thing one can do now is
to find a successful description of the mass matrix and look for
some symmetry principle behind it. The 
most successful full mass matrix description 
in describing quark mass and mixing, at least at a 
phenomenological level, 
is the approach initiated by Fritzsch [5] and a
number of its variants [6,7] (we call them phenomenological mass
matrix approaches), although the basic physics of the dictated
matrix is often not quite clear.

In our earlier paper we have presented a Fritzsch matrix type
model that describes the
case (i), in which the  $\nu_\tau-\nu_\mu$ mixing angle comes out to be
large when the $\nu_\mu-\nu_e$ mixing angle is small [8]. Indeed, the new 
Superkamiokande result together with a small angle solution of the
MSW scenario for the solar neutrino problem 
fits very well with the model, once one admits 
hierarchical massive neutrinos.  In this paper we focus on the
more heterodox possibility of the almost degenerate case, 
and discuss whether any
simple, natural-looking mass matrix exists that leads to this
unusual mass pattern together with a large mixing angle that explains
atmospheric neutrino oscillation. 
We consider that the large difference between
lepton and quark mixings should be ascribed to the Majorana character of the
neutrino. 

One of the most attractive description of the quark sector in
the phenomenological mass matrix approach starts with an
S$_3(R)\times$S$_3(L)$ symmetric mass term (often called ``democratic''
mass matrix)[6], and adds a small term that breaks
this symmetry [7], i.e.,
$$M_q= {K_q \over 3}
             \left[ \matrix{1 & 1 & 1 \cr
                            1 & 1 & 1 \cr
                            1 & 1 & 1 \cr
                                         } \right]
+\left[ \matrix{\delta_1^q & 0 & 0 \cr
                         0 & \delta_2^q & 0 \cr
                         0 & 0 & \delta_3^q \cr
                                              } \right]
,\eqno{(3)}$$
where $q=$ up and down, and quarks belong to 
{\bf 3}={\bf 2}$\oplus${\bf 1} of S$_3(L)$ or S$_3(R)$ as discussed in [6]. 
The first term is a unique representation of
the S$_3(R)\times$S$_3(L)$ symmetric matrix.
This matrix is diagonalised as
$$U_q^\dagger M_q U_q = {\rm diag}(m_1^q, m_2^q, m_3^q), \eqno{(4)}$$
where
$$m_1^q=(\delta_1^q+\delta_2^q+\delta_3^q)/3-\xi^q/6$$
$$m_2^q=(\delta_1^q+\delta_2^q+\delta_3^q)/3+\xi^q/6 \eqno{(5)}$$
$$m_3^q=K_q+(\delta_1^q+\delta_2^q+\delta_3^q)/3$$
with
$$\xi=[(2\delta_3^q-\delta_2^q-\delta_1^q)^2+3(\delta_2^q-\delta_1^q)^2
                                                ]^{1/2}, \eqno{(6)}$$
where terms of $O(\delta/K)$ are ignored.
The matrix that diagonalises $U_q=AB_q$ reads
$$A= \left[ \matrix{1/\sqrt 2 & 1/\sqrt 6 & 1/\sqrt 3 \cr
                   -1/\sqrt 2 & 1/\sqrt 6 & 1/\sqrt 3 \cr
                           0 & -2/\sqrt 6 & 1/\sqrt 3 \cr
                                         } \right], \eqno{(7)}$$

$$B_q=\left[ \matrix{ \cos \theta^q & - \sin \theta^q & 
                                          \lambda^q\sin 2\theta^q \cr
                  \sin \theta^q  & \cos \theta^q & 
                                        \lambda^q\cos 2\theta^q   \cr
              -\lambda^q\sin 2\theta^q    & \lambda^q\cos 2\theta^q & 1 \cr
                                         } \right], \eqno{(8)}$$
with
$$\tan 2\theta^q\simeq-\sqrt{3}{\delta_2^q-\delta_1^q \over 
  2\delta_3^q-\delta_2^q-\delta_1^q}, \eqno{(9)}$$ 
and $\lambda_q={1 \over \sqrt 2}{1 \over 3K_q}\xi_q$.
$A$ is the matrix that diagonalises the first term of (3). It has been
shown [7] 
 that all quark masses and mixing angles are successfully
given by taking $\delta_1=-\epsilon_q$, $\delta_2=\epsilon_q$ and
$\delta_3=\delta_q$ in (3), and adjusting these parameters in a way
$K_q\gg \delta_q > \epsilon_q$.

We assume the same structure for the charged leptons, and denote the
matrices with the script $\ell$ instead of $q$. Analogous to the
quark sector we choose $\delta_1^\ell=
-\epsilon_\ell$, $\delta_2^\ell=\epsilon_\ell$ and 
$\delta_3^\ell=\delta_\ell$. The
three mass eigenvalues (see (5)) are then 
$$m_1\simeq -\epsilon_\ell^2/2\delta_\ell, \hskip5mm m_2\simeq 2\delta_\ell/3+
\epsilon_\ell^2/2 \delta_\ell,
\hskip5mm m_3\simeq K_\ell+\delta_\ell/3, \eqno{(10)}$$
and the angle that appears in (8) is
$$\sin \theta_\ell \simeq- \sqrt{|m_1/m_2|} \  . \eqno{(11)}$$

Let us turn to the neutrino sector. Assuming that the neutrinos are of
the Majorana type, we have two invariant mass terms $\bf 2_L\times 2_L$ and
$\bf 1_L\times 1_L$.
Then, there  are two candidate matrices that are invariant under S$_3(L)$:
$$ \left[ \matrix{1 & 0 & 0 \cr
                  0 & 1 & 0 \cr
                  0 & 0 & 1 \cr} \right],
\hskip1cm
\left[ \matrix{0 & 1 & 1 \cr
                  1 & 0 & 1 \cr
                  1 & 1 & 0 \cr} \right]. \eqno{(12)}$$
Here we take the first form as the main mass term $M_\nu^{(0)}$ 
with a coefficient
$K_\nu$, deferring discussion about the second matrix later in this
paper. We then break symmetry by adding a small term with two adjustable
parameters. 
As a simple parametrisation we take 
$$M_\nu^{(1)}=\left[ \matrix{0 & \epsilon_\nu & 0 \cr
                 \epsilon_\nu  & 0 & 0 \cr
                  0 & 0 & \delta_\nu \cr} \right].\eqno{(13)}$$
An alternative natural choice to lift the mass degeneracy may be 
diag($-\epsilon_\nu, \epsilon_\nu, \delta_\nu)$, which we shall 
also discuss later.
The mass eigenvalues of $M_\nu=M_\nu^{(0)}+M_\nu^{(1)}$ 
are $K_\nu\pm\epsilon_\nu$, and 
$K_\nu+\delta_\nu$, and the matrix
that diagonalises $M_\nu$  ($U^TM_\nu U=$diagonal) is
$$U_\nu= \left[ \matrix{1/\sqrt 2 & 1/\sqrt 2 & 0 \cr
                   -1/\sqrt 2 & 1/\sqrt 2 & 0 \cr
                           0 & 0 & 1 \cr
                                         } \right]. \eqno{(14)}$$
 That is, our $M_\nu$ represents three degenerate neutrinos, with the degeneracy
lifted by a small parameters. In the literature [9] degenerate neutrinos 
are discussed starting with $M_\nu={\rm diag}(1,1,1)$ as an assumption. 
Our argument
provides a reason for degenerate neutrinos by treating quarks and leptons in
an equal-footing way. 

The lepton mixing angle (Cabibbo-Kobayashi-Maskawa matrix)
as defined by $V_\ell=(U_\ell)^\dagger U_\nu$ is thus given
by
$$V_\ell=(AB_\ell)^\dagger U_\nu\simeq
\left[ \matrix{1 & -(1/\sqrt 3)\sqrt{(m_e/m_\mu)} 
                                & (2/\sqrt 6) \sqrt{(m_e/m_\mu)} \cr
                    \sqrt{(m_e/m_\mu)} & 1/\sqrt 3 & -2/\sqrt 6 \cr
                           0 & 2/\sqrt 6 & 1/\sqrt 3 \cr
                                         } \right], \eqno{(15)}$$
where $m_1$ and $m_2$ in (10) are identified with $m_e$ and $m_\mu$. 
We note 
that the neutrino mass parameters  do not appear in this
mixing matrix.  The
parameters $K_\ell$, $\delta_\ell$ and $\epsilon_\ell$ are determined so that
the charged lepton analogue of
(5) gives electron, $\mu$ and $\tau$ masses for the charged lepton sector, and
$\epsilon_\nu K_\nu$ and $\delta_\nu K_\nu$ are fixed by the neutrino mass
difference explored by the oscillation effect:
$\Delta m_{32}^2=m_{\nu_3}^2-m_{\nu_2}^2\approx 0.5 \times 10^{-2}$ eV$^2$ 
[1] and  $\Delta m_{21}^2=m_{\nu_2}^2-m_{\nu_1}^2\approx 0.8 \times 
10^{-5}$ eV$^2$ [3]  
are obtained from the atmospheric and solar neutrino oscillation
(we take the small angle solution of the MSW scenario for the solar
neutrino problem [10]). The normalisation $K_\nu$ is not
fixed unless one of the neutrino masses is known, but it is not important
for our argument, since 
the lepton mixing matrix is almost 
independent of the details of these parameters except for the 
$m_e/m_\mu$ ratio, as we see in (15) where small terms are ignored.
If we retain all small terms, the lepton mixing angle is predicted
to be
$$V_\ell= \left[ \matrix{0.998 & -0.045 & 0.05 \cr
                        0.066 & 0.613 & -0.787 \cr
                        0.005 & 0.789 & 0.614 \cr
                                         } \right] \eqno{(16)}$$
instead of (15), where 
$K_\ell=1719$ MeV, $\delta_\ell=163$ MeV, $\epsilon_\ell=
11$ MeV, $\delta_\nu=0.0025$ eV and $\epsilon_\nu=2\times 10^{-6}$ eV are
used and $K_\nu=1$ eV is assumed (the matrix depends very little on the
assumption of  $K_\nu$).

$\nu_\mu-\nu_\tau$ oscillation is then given by
$$P(\nu_\mu\rightarrow\nu_\tau)\simeq 4V_{23}^2V_{33}^2\sin^2\bigg({\Delta
m_{32}^2
\over 4E}L\bigg)\simeq {8\over 9}\sin^2\bigg({\Delta m_{32}^2
\over 4E}L\bigg), \eqno{(17)}$$
which represents that mixing is  close to maximal. With 
a more accurate matrix (16) the factor 8/9 is
modified to 0.93. This means that the survival probability
of $\nu_\mu$ is 54\% for average neutrino oscillation, in very good
agreement with  the finding at the Superkamiokande [1]
(and also the result from Kamiokande [2]). 
For the $\nu_e-\nu_\mu$ oscillation
$\sin^22\theta\simeq 8\times 10^{-3}$, which also agrees with the neutrino 
mixing corresponding to the
small angle solution of the MSW scenario for the solar neutrino problem [3,4].

Let us now discuss constraints placed on this scenario. Since we have
assumed the Majorana type of neutrinos, we must require the condition
that the presence of the effective Yukawa term 
$${\cal L}={h \over M}\ell_L\ell_L H H \eqno{(18)}$$
($\ell_L$ is the left handed lepton doublet, $H$ the Higgs field,
$M$ an effective mass and $h$ is the Yukawa coupling)
should not erase baryon number of the universe above the weak mass scale [11]. 
Namely,
the condition reads
$$h^2/M^2<g/(M_{\rm planck}T) \eqno{(19)}$$
with $g$ the effective number of relativistic degrees of freedom
at temperature $T$ and $T$ is set equal to $10^{12}$ GeV [12], above which 
sphalerons do not
work to violate $B+L$. This yields
$$m_\nu < {h\over M}\langle H\rangle^2 \simeq 1 {\rm eV}. \eqno{(20)}$$
It is obvious that the scenario requires all neutrino masses to be
larger than $\approx$0.07 eV, the limit set by $\Delta m_{23}^2$ itself.

A very important constraint comes from neutrinoless double beta decay
experiments. The latest result on the lifetime of 
$^{76}$Ge$\rightarrow$$^{76}$Se,
$\tau_{1/2}>1.1\times 10^{25}$ yr [13] yields an upper limit on the Majorana
neutrino mass 0.4 eV [14] to 1.1 eV [15] depending on which nuclear model
is adopted for nuclear matrix elements (see [16] for a review of the
matrix element). This limit coincides with what is
derived from the survival of baryon number of the universe.
We are then left with quite a narrow window for the neutrino mass 
$0.1 {\rm eV}\leq m_{\nu_e}\simeq m_{\nu_\mu}\simeq m_{\nu_\tau}\leq 
1 {\rm eV}$
for the present scenario to be viable.
It will be most interesting to push down the lower limit of neutrinoless
double beta decay lifetime; if the limit on neutrino mass is lowered by
one order of magnitude the degenerate neutrino mass scenario 
as discussed in this paper will be ruled out. 

The argument we have made above is of course by no means unique, 
and a different
assumption on the matrix leads to a different mass-mixing relation. 
Let us briefly discuss the
consequence of the other matrices we have encountered in the line of our 
argument above. If we adopt the symmetry breaking term alternative to (13), 
$$M_\nu^{(1)}=\left[ \matrix{-\epsilon_\nu & 0 & 0 \cr
                 0  & \epsilon_\nu & 0 \cr
                  0 & 0 & \delta_\nu \cr} \right] \eqno{(21)}$$
in parallel to the charged lepton and quark sectors,
we obtain the lepton mixing matrix to be
$$V_\ell\simeq \left[ \matrix{1/\sqrt 2 & -1/\sqrt 2 & 0 \cr
                   1/\sqrt 6 & 1/\sqrt 6 &  -2/\sqrt 6 \cr
                           1/\sqrt 3 &  1/\sqrt 3 &  1/\sqrt 3 \cr
                                         } \right] . \eqno{(22)}$$
This is identical to the matrix presented by
Fritzsch and Xing [17], where they {\it assumed} the
neutrino mass matrix basically identical to the case discussed here.
  For this case
we obtain 
$$\sin^22\theta_{12}\simeq 1,\hskip5mm \sin^22\theta_{23}\simeq 8/9.
                  \eqno{(23)}$$
The  maximal mixing
is derived for the (1,2) sector, whereas the 
mixing angle for the (2,3) sector is unchanged, again 
irrespective of the details of  
neutrino masses. Namely, this case can accommodate the ``just-so''
scenario for the solar neutrino problem due to neutrino 
oscillation in vacuum [4],
instead of the small angle solution of the MSW scenario. The constraints
from double beta decay, baryon number of the universe etc. discussed above
all apply to this case in the same way.

There is another branch of the argument within our framework. If the
second form is adopted for $M_\nu^{(0)}$ in (12), we are led to
small mixing angles for all neutrinos. Therefore, the choice of a 
diagonal form in (12) was crucial to obtain a large mixing angle
for the lepton sector.  We do not discuss this case further here. 

In this paper we have shown that there exist simple lepton mass matrices
derived from some symmetry principle with a simple breaking term,  
which gives rise to almost degenerate neutrinos with
the (2,3) component almost maximally mixed. Our lepton matrix also
gives mixing angle for the (1,2) sector consistent with either small
angle solution of the MSW neutrino conversion scenario or
maximal mixing solution included in the ``just-so'' scenario of
neutrino oscillation in vacuum, as required from the solar neutrino
problem. 
The prediction we discussed for double beta decay is interesting,
 but does not depend on our specific model.
The allowed window of the neutrino mass in our scenario is
very narrow: this motivates us to push hard double beta decay
experiment to set a more stringent limit on the Majorana neutrino
mass.

\vskip 30 mm
\noindent
{\large\bf Acknowledgements}

We should like to thank Yoichiro Suzuki and Yoji Totsuka for stimulating
discussions.

\vskip 1.5 cm
\noindent
{\large \bf References}\par
\vskip 0.3 cm
\noindent
 [1] Y. Totsuka, in  Talk given at 18th International Symposium  on     
  Lepton-Photon \par Interactions, July 28 - August 1, 1997, Hamburg.
  to be published in the Proceedings.    

\noindent
 [2] K. S. Hirata et al. Phys. Lett. B {\bf 280}, 146 (1992);\par
        R. Becker-Szendy et al. Phys. Rev. D {\bf 46}, 3720 (1992);\par
        SOUDAN2 Collaboration, W. W. M. Allison et al., Phys. Lett. B {\bf
391}, 491 (1997).

\noindent
 [3] E.g., J. N. Bahcall and P. I. Krastev, Phys. Rev. D {\bf 53}, 4211 (1996).

\noindent
 [4] E.g., V. Barger, R. J. N. Phillips and K. Whisnant, Phys. Rev.
        Lett. {\bf 69}, 3135 (1992).

\noindent
 [5] H. Fritzsch, Phys. Lett. B {\bf 70}, 436 (1977); Nucl. Phys. B {\bf
155}, 189 (1979).

\noindent
 [6] H. Harari, H. Haut and J. Weyers, Phys. Lett. B {\bf 78}, 459 (1978).

\noindent
 [7]  Y. Koide,  Phys. Rev. D {\bf 28}, 252(1983); {\bf 39}, 1391 (1989).

\noindent
 [8] M. Fukugita, M. Tanimoto and  T. Yanagida, Prog. Theor. Phys. {\bf 89},
        263 (1993).
		
\noindent
 [9]  D. O. Caldwell and R.N. Mohapatra, Phys. Rev. D {\bf 48} 3259 (1993); 
 
     Phys. Rev. D {\bf 50} 3477 (1994);
	 
     S. T. Petcov and A. Yu. Smirnov,  Phys. Lett. B {\bf 322} 109 (1994);

\noindent		
 [10] We have started with assuming that the solar neutrino problem is 
      explained by 
	  
	   neutrino oscillation.  Hence, we do not consider the LSND 
      experiment (C. 
	  
	   Athanassopoulos et al., Phys. Rev. Lett. {\bf
      77} 3082 (1996)) which is not compatible 
	  
	   with this view.

 \noindent
 [11] M. Fukugita and T. Yanagida, Phys. Rev. D {\bf 42}, 1285 (1990).
 
\noindent
 [12] J. Ambj\o rn, T. Askgaard, H. Porter and M. E. Shaposhnikov,\par
        Nucl. Phys. B {\bf 353}, 346 (1991).
       
 \noindent
 [13] H. V. Klapdor-Kleingrothaus, 
   in  Proceedings of the 17th International
   Conference
   
    on Neutrino Physics and Astrophysics, Helsinki, Finland,
   ed.  K. Enqvist et al.,
   
    World Scientific, 317 (1996).

 \noindent
 [14] T. Tomoda, A. Faessler, K. W. Schmid and F. Gr\"ummer,
        Nucl. Phys. A {\bf 452},\par
		 591 (1986).

\noindent
 [15] J. Engel, P. Vogel, X. Ji and S. Pittel, Phys. Lett. B {\bf 225}, 5
(1989).

 \noindent
 [16] M. Fukugita and T. Yanagida, in Physics and Astrophysics of
         Neutrinos, \par
	ed. by M. Fukugita and A. Suzuki (Springer, Tokyo, 1994). 

 \noindent
 [17] H. Fritzsch and Z. Xing, Phys. Lett. B {\bf 372}, 265 (1996).

\end{document}